\documentclass[twocolumn,showpacs,preprintnumbers,amsmath,amssymb]{revtex4}

\usepackage{xspace}
\usepackage{amsfonts}
\usepackage{graphicx}
\usepackage{url}
\usepackage[dvipdfm,%
              colorlinks,%
          linkcolor=blue,%
        citecolor = blue,%
              hyperindex,%
        plainpages=false,%
         urlcolor = blue,%
       pdfstartview=FitH]{hyperref}

\begin{document}

%\preprint{}

\title{\textsf{Carrier Statistics and Quantum Capacitance of Graphene Sheets and Ribbons }}

\author{Tian Fang}
\author{Aniruddha Konar}
\author{Huili Xing}
\author{Debdeep Jena}
\email[Electronic mail: ]{djena@nd.edu}
%\author{Aniruddha Konar}
%\email[Electronic mail: ]{ycao1@nd.edu}

\affiliation{Department of Electrical Engineering, University of Notre Dame, IN, 46556, USA}

\date{\today}

\begin{abstract}
In this work, fundamental results for carrier statistics in graphene 2-dimensional sheets and nanoscale ribbons are derived.  Though the behavior of intrinsic carrier densities in 2d graphene sheets is found to differ drastically from traditional semiconductors, very narrow (sub-10 nm) ribbons are found to be similar to traditional narrow-gap semiconductors.  The quantum capacitance, an important parameter in the electrostatic design of devices, is derived for both 2d graphene sheets and nanoribbons. 
\end{abstract}



\maketitle
%%%%% Text of paper:
%%

%=================================================================

%-------------Begin Intro------------------->>

Graphene, a 2-dimensional (2D) honeycomb structure of carbon atoms, has generated intense interest recently \cite{geim07natmat, novo05nat, prl07kim, science06gatech}.  It has been now demonstrated that narrow graphene nanoscale ribbons (GNRs) posses bandgaps that are tuned by the ribbon width \cite{prl07kim}.  These properties, along with the good transport properties of carriers (high mobility, high Fermi velocity) suggest that it is possible that graphene may be used in the near future in high speed electronic devices \cite{apl06gnrIntel, edl07grfet}.  In spite of rapid advances in the study of transport properties of graphene, basic tools of semiconductor device design such as temperature dependent carrier statistics and electrostatic properties such as quantum capacitance remain unexplored.  This work investigates these properties for both 2D sheets, and GNRs in a comparative fashion, and analytical results for these quantities are presented.

%-------------End Intro------------------->>
%-------------Begin 2d graphene carrier stats------------------->>
The dispersion of mobile $\pi$-electrons in graphene in the 1st Brillouin Zone (BZ) is given by 
\begin{equation}
E({\bf k}) = s \hbar v_{F} |{\bf k}|,
\label{dispersion}
\end{equation}
where $s= +1$ is the conduction band (CB), and $s= -1$ is the valence band (VB), $\hbar$ is the reduced Planck's constant, $v_{F} \sim 10^{8}$ cm/s is the Fermi velocity of carriers in graphene, and $|{\bf k}| = \sqrt{k_{x}^{2} + k_{y}^{2}}$ is the wavevector of carriers in the 2D ($x-y$) plane of the graphene sheet \cite{science06gatech, rmp07cnt}.  The point $|{\bf k }|= 0$, referred to as the `Dirac point', is a convenient choice for the reference of energy, thus $E(|{\bf k}| = 0) = $ 0 eV.  Each ${\bf k}-$point is two-fold spin degenerate ($g_{s} = 2$), and there are two valleys in the 1st BZ (the $K$ \& $K^{\prime}$ valleys), $g_{v} = 2$.  

In an undoped layer of graphene in thermal equilibrium, there are mobile electrons in the CB and holes in the VB, similar to the intrinsic carriers in a pure bulk semiconductor.  To find the 2D sheet density of such intrinsic carriers in graphene, the linear density of states (DOS) 
\begin{equation}
\rho_{gr}(E) = \frac{ g_{s} g_{v} }{ 2 \pi (\hbar v_{F})^{2} } |E|,
\end{equation}
is used to write the 2D electron gas sheet density in graphene as
\begin{equation}
n = \int_{0}^{\infty} dE \rho_{gr}(E) f(E),
\label{ngeneral}
\end{equation}
where $f(E)$ is the Fermi-Dirac distribution function given by $f(E) = (1 + \exp{[ (E - E_{F})/ k T ]}  )^{-1}$,  $k$ being the Boltzmann constant, $T$ the absolute temperature, and $E_{F}$ is the Fermi-level.  With the aid of the dimensionless variables $u = E / kT$ and $\eta = E_{F}/kT$ (s is the index for conduction/valence bands), the electron density may be rewritten as
\begin{equation}
n = \frac{ 2 }{ \pi} (\frac{kT}{\hbar v_{F}})^{2} \Im_{1}(+ \eta) ,
\label{nparticular}
\end{equation}
and the hole density is symmetric, given by 
\begin{equation}
p = \frac{ 2 }{ \pi} (\frac{kT}{\hbar v_{F}})^{2} \Im_{1}( - \eta) ,
\end{equation}
where $\Im_{j}(\eta) = 1/\Gamma(j+1) \int_{0}^{\infty} du \cdot u^{j}/(1+ e^{u-\eta})$ is the Fermi-Dirac integral with $j=1$.   

Under thermal equilibrium and under no external perturbation (no applied bias, no optical illumination), the Fermi level is unique, and moreover, it is exactly at the Dirac point ($E_{F} = 0$ eV).  Then, the intrinsic carrier concentration in 2D graphene is given by
\begin{equation}
n=p= n_{i} = \frac{\pi}{6} (\frac{ k T }{ \hbar v_{F} })^{2},
\end{equation}
which is dependent on only one material parameter - the Fermi velocity.  The point to note is that the intrinsic sheet density of electrons/holes does not depend on temperature exponentially; it has a $T^{2}$ dependence, due to the absence of a   bandgap, and the linear energy dispersion.  At room temperature, the intrinsic electron and hole sheet densities evaluate to $n_{i} \sim 9 \times 10^{10}$ cm$^{-2}$.  
%----------------------------------------------------------------
%Figure
\begin{figure}%[h]
\begin{center}
\leavevmode 
\includegraphics[width=3in]{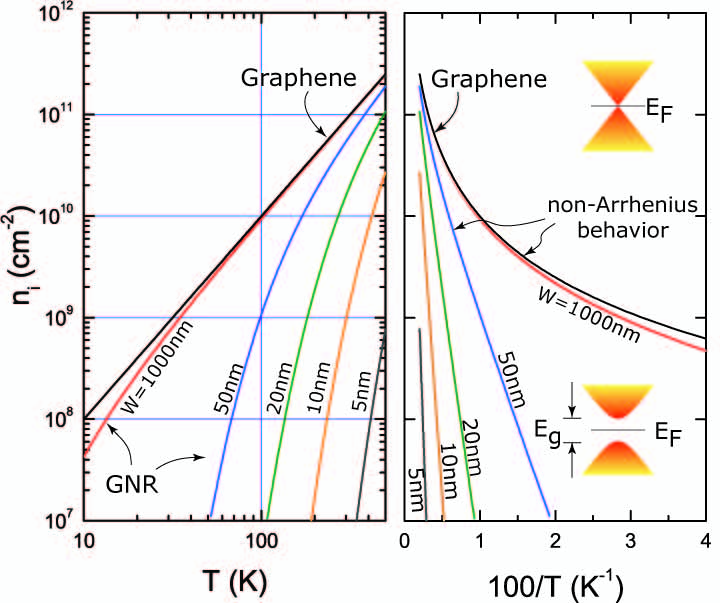}
\caption{Intrinsic sheet carrier concentrations in graphene sheets and nanoribbons.  Wide GNRs and 2d graphene have non-Arrhenius dependence on temperature, which becomes increasingly Arrhenius-like as the ribbon width decreases.} 
\label{fig1}
\end{center}
\end{figure}
%Figure
%----------------------------------------------------------------
%-------------End 2d graphene carrier stats------------------->>
%-------------Begin GNR carrier stats------------------->>

The situation changes for nanoscale ribbons cut from infinite graphene sheets.  Consider a GNR of width $W$.  Current experimental evidence suggests no clear dependence of the bandgap of GNRs on the chirality \cite{prl07kim}.  Regardless, the results derived here remain applicable for GNRs with bandgaps.  We make the assumption that the electron and hole quasimomenta are isotropic in the graphene plane. By aligning the $x-$axis along the longitudinal direction of the ribbon, the electron wavevector in the $y-$direction is quantized by hard-wall boundary conditions to be $k_{y} = n \pi / W$ ($n = \pm1, \pm2, ...$), and the energy dispersion relation (Equation \ref{dispersion}) for the $n^{th}$ subband becomes 
\begin{equation}
E(n,k_{x}) = s \hbar v_{F} \sqrt{ k_{x}^{2} + (\frac{n \pi}{W})^{2} },
\label{dispersion_GNR}
\end{equation}
indicating that the CB ($n>0$) \& VB ($n<0$) split into a number of 1D subbands, indexed by $n$.  It is obvious that  breaking the symmetry of a graphene sheet by cutting out a ribbon opens up a band gap.  For the isotropic case assumed here, the bandgap for a GNR of width $W$ is given by 
$E_{g} = E(+1, 0) - E( -1,  0) = 2 \pi \hbar v_{F} / W$, dependent only on the Fermi velocity and the width of the GNR.  The DOS relation for the $n^{th}$ 1D subband is then given by
\begin{equation}
\rho_{GNR}( n, E ) = \frac{4}{\pi \hbar v_{F} } \cdot \frac{ E }{ \sqrt{ E^{2} - E_{n}^{2} } } \Theta (E - E_{n}),
\label{dos_gnr}
\end{equation}
where $\Theta(...)$ is the Heaviside unit step function, and $E_{n} = n \pi \hbar v_{F}/ W = n E_{g}/2$.  This directly leads to a {\em total} DOS $\rho_{GNR}(E) = \sum_{n} \rho_{GNR}( n, E)$.  The expression for the total DOS is the same for the CB and VB, and exhibits van-Hove singularities at energies $E_{n}$ from the Dirac point.  The electron density as a result is given by
\begin{equation}
n = \frac{4kT}{\pi \hbar v_{F}} \sum_{n>0} S ( x_{n}, \eta ) ,
\label{nGNRexact}
\end{equation}
where $x_{n} = E_{n} / k T$, $\eta = E_{F}/kT$, and
\begin{equation}
S ( x, \eta) = \int_{x}^{\infty} \frac{u}{\sqrt{ u^{2} - x^{2} }} \frac{du}{1+e^{ u - \eta }}.
\end{equation}
%----------------------------------------------------------------
%Figure
\begin{figure}%[h]
\begin{center}
\leavevmode 
\includegraphics[width=3in]{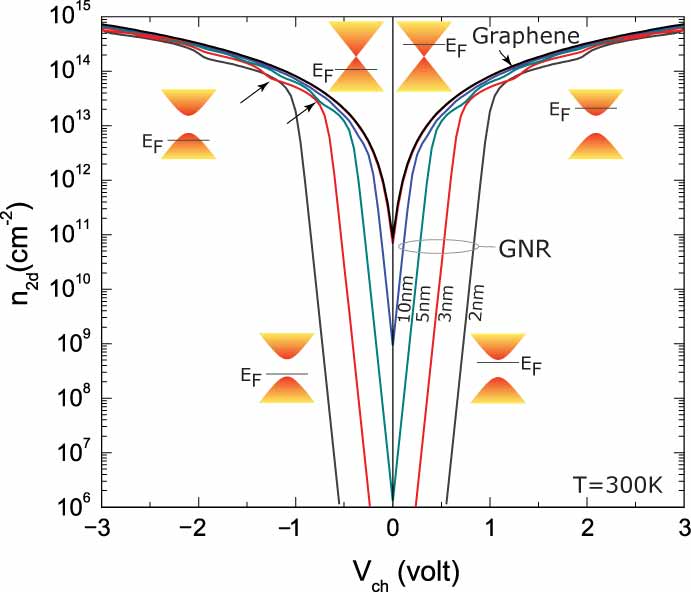}
\caption{2D carrier concentration in graphene and GNRs of different widths as a function of the location of the Fermi level.  Though narrow GNRs exhibit large charge modulation due to the existence of a gap, they become similar to 2D graphene sheets when the Fermi level is deep inside the bands.} 
\label{fig2}
\end{center}
\end{figure}
%Figure
%----------------------------------------------------------------
The intrinsic carrier concentration in GNRs is obtained when $\eta = E_{F}=0$, i.e., the Fermi level is at mid-gap; this leads to $n_{i} = (4kT/ \pi \hbar v_{F}) \sum_{n} S(x_{n},0) $.  For narrow GNRs, $E_{g} \gg kT$, then one can use the approximation $S(x,0) \approx xK_{1}(x) $ where $K_{1}(...)$ is the Bessel function of 1st order, and the asymptotic approximation of the Bessel function $K_{1}(x) \approx \sqrt{\pi / 2 x} \exp{(-x)}$ for large $x$ to write the intrinsic carrier density of GNRs as
\begin{equation}
n_{i}  \approx \frac{4}{W} \sqrt{ \frac{\pi k T}{E_{g}} } \sum_{n} \sqrt{n} e^{-n \cdot \frac{ E_{g} }{ 2kT }}.
\end{equation}
For bandgaps well in excess of the thermal energy, it suffices to retain only the first term in the summation to recover the familiar dependence
\begin{equation}
n_{i} \approx \frac{4}{W} \sqrt{ \frac{\pi k T}{E_{g}} } e^{-\frac{E_{g}}{2kT}}.
\end{equation}
This relation has to be used with caution when experimentally extracting bandgaps from the slope of Arrhenius-like plots; it is applicable {\em only} when the bandgap is well in excess of the thermal energy.  The 1D carrier concentration of GNRs may be converted to an effective 2D sheet density by writing $n_{2d} = n_{1D}/W$ for comparing their properties with graphene, as is done in Figure \ref{fig1}.  This figure shows that the intrinsic carrier concentrations in GNRs differs significantly from graphene only if the ribbon widths are below $\sim$0.1 $\mu m$, and indicates when Arrhenius dependence of intrinsic carrier concentrations is valid.
%-------------End GNR carrier stats------------------->>
%-------------Begin graphene n vs vch------------------->>

The carrier sheet density in graphene can be changed by an electrostatic gate voltage, and the on-state sheet densities can approach, and exceed those in conventional field-effect transistors (FETs).  If the Fermi level in a 2D graphene sheet is driven from the Dirac point to $E_{F} = \eta kT$ electrostatically by means of a gate voltage, then the electron density is given by $n = n_{i} \Im_{1}(\eta) / \Im_{1}(0)$, and the hole density by $p = n_{i} \Im_{1}(-\eta) / \Im_{1}(0) $, leading to a mass-action law $n p = n_{i}^{2} \Im_{1}(\eta) \Im_{1}(-\eta) / \Im_{1}^{2}(0)$.
%-------------End graphene n vs vch------------------->>
%-------------Begin GNR n vs vch------------------->>
%----------------------------------------------------------------
%Figure
\begin{figure}%[h]
\begin{center}
\leavevmode 
\includegraphics[width=3in]{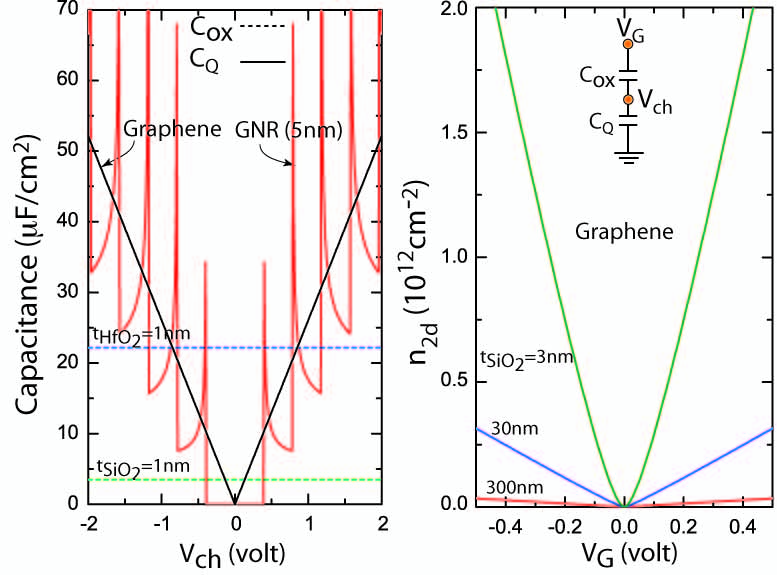}
\caption{Left: Quantum capacitance of 2D graphene and a 5nm GNR compared with the parallel-plate capacitance of 1 nm SiO$_{2}$ \& HfO$_{2}$. Right: 2D carrier density in a graphene sheet as a function of gate voltage for different oxide thicknesses.} 
\label{fig3}
\end{center}
\end{figure}
%Figure
%----------------------------------------------------------------
Similarly, if the local electrostatic potential in a GNR is tuned by a gate voltage such that the Fermi level is at $E_{F} = \eta kT$, then the electron density is given by equation \ref{nGNRexact}.  For $\eta > x \gg 1$, one can make the approximation $ (1+\exp{( u - \eta )})^{-1} \approx \Theta( \eta - u ) $ to rewrite the 1D electron concentration as 
\begin{equation}
n \approx \frac{4}{\pi \hbar v_{F}} \sum_{n} \sqrt{ E_{F}^{2} - E_{n}^{2}  } \Theta (E_{F} - E_{n} ),
\end{equation}
On the other hand, for a non-degenerate condition when the Fermi level is located inside the GNR bandgap, using the approximation $S(x,\eta) \approx \sqrt{ \pi x / 2 } \exp(\eta - x) $ the electron concentration may be written as $n \approx n_{i} e^{\eta}$, and similarly, for holes, $p \approx n_{i} e^{-\eta}$, which is similar to traditional semiconductors.  Figure \ref{fig2} shows the calculated exact 2D carrier concentrations in graphene and GNRs of different widths as a function of the location of the Fermi level ($qV_{ch} = E_{F}$) at room temperature.  Though narrow GNRs exhibit large charge modulation due to the existence of a gap, they become similar to 2D graphene sheets when the Fermi level is deep inside the bands.  Ripples appear in the GNR density due to van-Hove singularities, as indicated by arrows.
%-------------End GNR n vs vch------------------->>
%-------------Begin graphene quantum capacitance------------------->>

An important quantity in the design of nanoscale devices is the quantum capacitance \cite{jap04pulfreyQuantCap}.  Writing the total charge in a graphene sheet with a local channel electrostatic potential $V_{ch}$ as $Q= q (p-n)$ where $q$ is the electron charge, and using the definition of quantum capacitance $C_{Q} = \partial Q / \partial V_{ch}$, one obtains for 2D graphene
\begin{equation}
C_{Q} = \frac{2q^{2}kT}{\pi (\hbar v_{F})^{2} } \ln{[2(1+\cosh{\frac{qV_{ch}}{kT}})]},
\end{equation}
which under the condition $qV_{ch} \gg kT$ reduces to 
\begin{equation}
C_{Q} \approx q^{2} \frac{2}{\pi} \frac{qV_{ch}}{(\hbar v_{F})^{2}} = q^{2} \rho_{gr}(qV_{ch}).
\end{equation}
If the electrostatic capacitance formed between a gate electrode and the graphene layer is given by $C_{ox} = \epsilon_{ox} / t_{ox}$, then the electron density in the graphene layer can be written as a function of the gate voltage as
\begin{equation}
n = n_{G} -  n_{Q}( \sqrt{1+ 2 \frac{n_{G}}{n_{Q}}} - 1 ),
\label{n2dVg}
\end{equation}
where $n_{G} = C_{ox}V_{G}/q$ is the traditional carrier density one would obtain by neglecting the quantum capacitance, and $n_{Q} = \frac{\pi}{2}(C_{ox} \hbar v_{F}/q^{2})^{2}$, which arises solely due to the quantum capacitance.  
%------------End graphene quantum capacitance------------------->>
%-------------Begin GNR quantum capacitance------------------->>
Applying the same method to find the quantum capacitance (per unit width) of GNRs, one obtains for the condition $\eta > x \gg 1$,
\begin{equation}
C_{Q} \approx \frac{4q^{2}}{\pi \hbar v_{F}} \sum_{n} \frac{\eta}{\sqrt{ \eta^{2} - x_{n}^{2} }} \Theta(\eta - x_{n} ) = q^{2} \rho_{GNR}(\eta).
\end{equation}
The quantum capacitance of 2D graphene and GNR is plotted in Figure \ref{fig3} (left), and compared with the oxide gate capacitance of 1 nm SiO$_{2}$ \& HfO$_{2}$.  Figure \ref{fig3} (right) shows the carrier density dependence in 2D graphene on the gate voltage (Eq. \ref{n2dVg}) for different SiO$_{2}$ thicknesses.  Gate modulation of the charge is strong but non-linear for very thin $t_{ox}$ since $C_{Q} \approx C_{ox}$ under that condition. The field-effect becomes weak but increasingly linear as $t_{ox}$ is increased since $C_{ox} << C_{Q}$ under that condition.

The results presented here would prove useful for the design of electronic devices using graphene sheets and GNRs.  The authors would like to thank G. Snider for useful discussions, and one of the authors (D.J.) thanks an NSF CAREER award for financial support.

%----------------------------------------------------------------
%Figure
%\begin{figure}%[h]
%\begin{center}
%\leavevmode \epsfxsize=3in \epsfbox{Figures/SLsXRD.eps}
%\caption{Measured and calculated (002) GaN X-ray diffraction spectrum for a
%9-period AlN/GaN superlattice.} \label{fig:SLsXRD}
%\end{center}
%\end{figure}
%Figure
%----------------------------------------------------------------

\bibliographystyle{unsrt}
%-------REFERENCES------------->
%\pagebreak

%FILL IN THE REFERENCES HERE ------->
%START------->
%\bibliography{biblio}

\pagebreak

\end{document}